\documentclass[prl,twocolumn]{revtex4}

\usepackage{ulem}
\usepackage{euscript}
\usepackage{epsfig}
\usepackage{amssymb}

\begin{document}

\title[ ]{Ultracold Fermion Cooling Cycle using Heteronuclear Feshbach Resonances}

\author{M. A. Morales\footnote{Present address: Department of Physics, University of Illinois at Urbana-Champaign, Illinois 61801.}}
\affiliation{Electron and Optical Physics Division, National Institute of Standards and Technology, Gaithersburg, Maryland 20899-8410}
\author{N. Nygaard}
\affiliation{Electron and Optical Physics Division, National Institute of Standards and Technology, Gaithersburg, Maryland 20899-8410}
\author{J. E. Williams}
\affiliation{Electron and Optical Physics Division, National Institute of Standards and Technology, Gaithersburg, Maryland 20899-8410}
\author{Charles W. Clark}
\affiliation{Electron and Optical Physics Division, National Institute of Standards and Technology, Gaithersburg, Maryland 20899-8410}

\begin{abstract}
We consider an ideal gas of Bose and Fermi atoms in a harmonic trap, with a Feshbach resonance in the interspecies atomic scattering that can lead to formation of fermionic molecules. We map out the phase diagram for this three-component mixture in chemical and thermal equilibrium. Considering adiabatic association and dissociation of the molecules, we identify a possible cooling cycle, which in ideal circumstances can yield an exponential increase of the phase-space density. 
\end{abstract}

\maketitle

{\bf{Introduction}}--Feshbach resonances~\cite{FeshbachTheory} have become a powerful tool for controlling atomic interactions in ultracold gases~\cite{Inouye1998a,Solitons,Collapse,BCS_BEC,BEC2molecules,Fermi2molecules,Stan2004a,Inouye2004a}. By tuning an external magnetic field, the energy of a molecular bound state can be adjusted to be on resonance with a scattering state of two free atoms, thus facilitating external control of the atomic interactions.
Tuning the molecular state below threshold can lead to the production of stable ultracold diatomic molecules. This has been accomplished both in bosonic~\cite{BEC2molecules} and fermionic gases~\cite{Fermi2molecules}. In all of these cases the diatomic molecules are homonuclear bosons. Two recent experiments, at MIT~\cite{Stan2004a} and at JILA~\cite{Inouye2004a} have observed Feshbach resonances in Bose-Fermi atomic mixtures. The molecular species in this case is a heteronuclear fermion. 

In this work we construct the phase diagram for an ideal trapped gas of bosonic and fermionic atoms, and the heteronuclear dimer formed from them~\cite{Yabu_Naboo}. In contrast to previous work on Fermi gases~\cite{Carr2004,Williams2004b}, we find that adiabatic passage from free atoms to molecules can lead to cooling of the system under conditions comparable to those of current experiments. Based on this observation we propose a thermodynamic cooling cycle, consisting of adiabatic interconversion between atoms and molecules in conjunction with selective removal of atoms. This cycle is capable of reducing the absolute temperature of the system while increasing the atomic phase-space density. 

{\bf{Equilibrium theory}}--We consider an ideal gas mixture composed of fermionic
$(f)$ and bosonic $(b)$ atoms and heteronuclear
fermionic molecules $(bf)$ trapped in an anisotropic
harmonic optical trap. The trapping frequencies for the different
components are taken to be:
\begin{equation}
\omega_{f}^{i} = \sqrt{\frac{\alpha_{f}^{i}}{m_{f}}}, \ \ \
\omega_{b}^{i} = \sqrt{\frac{\alpha_{b}^{i}}{m_{b}}}, \ \ \
\omega_{bf}^{i} =
\sqrt{\frac{\alpha_{f}^{i}+\alpha_{b}^{i}}{m_{f}+m_{b}}},
\label{frequencies}
\end{equation}
where $i=\lbrace x,y,z \rbrace$, $\alpha^i_k$ is proportional to the atomic polarizability, and
$m_k$ are the particle masses, with $k=\lbrace f,b,bf \rbrace$. The expression for the molecular frequency $\omega^i_{bf}$ in terms of the atomic frequencies is appropriate when the molecular internuclear separation is large compared to the mean atomic radii.  It is possible to change the ratio $\omega_{f}^{i}/\omega_{b}^{i}$ over a wide range by varying the wavelength of the trapping laser, and Ref. \cite{SC} gives specific prescriptions for doing this.  We assume chemical and thermal equilibrium: $\mu_{f} +
\mu_{b} = \mu_{bf}$ and $T_{f}=T_{b}=T_{bf}=T$, where $\mu_{k}$ are the chemical potentials of the three components, and $T$ is the temperature.
The energy of the molecular state as measured from the atomic dissociation continuum is $\epsilon_{\rm{res}}$, which is a function of an applied magnetic field. The results 
of this paper are based on considering the phase diagram of this three-component system as a function of $T$, $\epsilon_{\rm{res}}$, and the atomic populations. 

We employ a semi-classical approximation, such that populations of each component are given by 
\begin{eqnarray}
N_f(T,\mu_f) &=&  \left( \frac{k_{\rm{B}} T}{\hbar \bar{\omega}_f} \right)^3 \mathcal F_3(z_f), \\
\tilde N_b(T,\mu_b) &=& 
\left( \frac{k_{\rm{B}} T}{\hbar \bar{\omega}_b} \right)^3 \mathcal G_3(z_b), \\   
N_{bf}(T,\mu_{bf},\epsilon_{\rm{res}}) &=& \left( \frac{k_{\rm{B}} T}{\hbar \bar{\omega}_{bf}} \right)^3 \mathcal F_3(z_{bf}), 
\end{eqnarray}
where the fugacities $z_k$ are
\begin{equation}
z_f=  e^{\mu_f/k_{\rm{B}}T},\ \ 
z_b= e^{\mu_b/k_{\rm{B}}T},\ \
z_{bf}= e^{(\mu_{bf} -
\epsilon_{\rm{res}})/k_{\rm{B}}T}.
\end{equation}
Here $k_{\rm{B}}$ is Boltzmann's constant, and $\mathcal F_n(z)$ and $\mathcal G_n(z)$ are the Fermi-Dirac and Bose-Einstein integrals, respectively~\cite{Pathria,Williams2004b} .
This approximation is valid provided $k_{\rm{B}}T \gg \hbar\bar{\omega}_k$, where $\bar{\omega}_k= (\omega^x_k\omega^y_k\omega^z_k)^{\frac{1}{3}}$ is
the mean harmonic frequency.  $\tilde N_b(T,\mu_b)$
represents the number of non-condensed Bose atoms; we denote the number of Bose condensed atoms by $N_{c}$, and $N_b=N_{c}+\tilde{N}_b$.   

For a given $T$, $\epsilon_{\rm{res}}$, total number of atoms $N$, and boson-fermion population difference $\Delta N$ the atomic chemical potentials are determined by the conditions
\begin{eqnarray}
\label{totN} N &=& N_f(T,\mu_f)+
N_b(T,\mu_b)+2N_{bf}(T,\mu_{bf},\epsilon_{\rm{res}}), \\
\label{delNi} \Delta N &=& N_f(T,\mu_f)-N_b(T,\mu_b).
\end{eqnarray}

For $T$ above a critical temperature $T_c$ all bosons are thermal and $N_{c}=0$, whereas for $T<T_c$ a Bose-Einstein condensate (BEC) emerges: $N_{c}>0$, $\mu_b=0$ ($\mu_{bf}=\mu_f$), and 
\begin{equation}
\label{NbBelowTC} \tilde N_b(T,\mu_b=0)=\zeta(3) \left(\frac{k_{\rm{B}}T}{\hbar \bar{\omega}_b} \right)^3,
\end{equation}
where $\zeta(n)$ is the Riemann zeta function.

\begin{figure}
\centering
\includegraphics[width=\columnwidth]{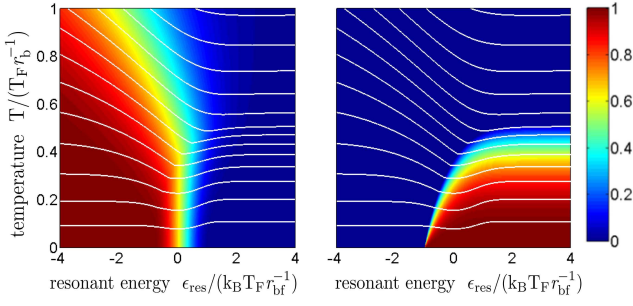}
\caption{Phase diagram for trapping parameters of the JILA system~\cite{JILA_parameters}, with $\Delta N = 0$; this representation is universal in $N$ subject to validity of the semi-classical approximation.  Left frame: molecule fraction, $\eta_{bf}$, as a function of $\epsilon_{res}$ and $T$; right frame: condensate fraction, $\eta_c$.  White lines indicate adiabatic trajectories, i.e. curves of constant entropy.} 
\label{PhaseDiagram}
\end{figure}

{\bf{Phase diagrams}}--We now discuss the phase diagrams resulting from the numerical
solutions to Eqs. (\ref{totN}) and (\ref{delNi}). Defining the molecular and condensate fraction as
$\eta_{bf} \equiv 2N_{bf}/N$ and
$\eta_{c} \equiv N_{c}/N_b$,
respectively, we plot these quantities in Fig.~\ref{PhaseDiagram} versus temperature and $\epsilon_{\rm{res}}$ for a system where $\Delta N=0$, using trapping parameters for the JILA  experiment~\cite{JILA_parameters}.
The $T$ and $\epsilon_{\rm{res}}$ axes are normalized by $T_{\rm{F}}/r_b$ and $k_{\rm{B}}T_{\rm{F}}/r_{bf}$, respectively,
where $T_{\rm{F}}$ is the Fermi temperature of the fermionic atoms,
$k_{\rm{B}}T_{\rm{F}}=\hbar\bar{\omega}_f(6xN)^{1/3}$, $x=(N_f+N_{bf})/N$ is the total fermionic fraction, $r_{bf}=\bar{\omega}_f/\bar{\omega}_{bf}$ and $r_b=\bar{\omega}_f/\bar{\omega}_b$. For a finite population difference the maximum number of molecules is $N_{bf}^{\rm{max}}=(N-|\Delta N|)/2$.

When $\Delta N < 0$, there will always
be free bosons in the system, and condensation is possible for any
value of $\epsilon_{\rm{res}}$.  When $\Delta N \geq 0$, there is a critical resonance energy, $\epsilon^{(c)}_{\rm{res}}$, at which the condensate fraction vanishes at $T=0$. This can be seen in Fig. \ref{PhaseDiagram}.
Setting $N_b=0$ in Eqs. (\ref{totN}) and (\ref{delNi}) and $\mu_{bf}=\epsilon^{(c)}_{\rm{res}}+(6N_{bf})^{1/3}\hbar\bar{\omega}_{bf}$ at $T=0$ we find that
\begin{equation}
\frac{\epsilon_{\rm{res}}^{(c)}}{k_{\rm{B}}T_{\rm{F}}}=\left( 2-\frac{1}{x} \right)^{1/3} - \frac{1}{r_{bf}} \left( \frac{1}{x}-1 \right)^{1/3}.
\end{equation}
In a similar manner we can solve for the critical temperature $T_c$ for BEC in the limit $\epsilon_{\rm{res}}/k_{\rm{B}}T_{\rm{F}}
\gg 1$, where $N_{bf}=0$. 
From Eq. (\ref{NbBelowTC}) we find
\begin{equation}
\frac{T_c}{T_{\rm{F}}}=[6\zeta(3)]^{-1/3} \frac{1}{r_b} \left( \frac{1}{x}-1 \right)^{1/3},
\end{equation}
In the case of equal populations $x=1/2$, the limits
take a simple form 
$\epsilon_{\rm{res}}^{(c)} =-k_{\rm{B}}T_{\rm{F}}/r_{bf}$ and
$T_c  \simeq 0.518 T_{\rm{F}}/r_b$.
These two points define the left and top boundaries of the condensate region in Fig.~\ref{PhaseDiagram}.

{\bf{Adiabatic sweep}}--
During an adiabatic sweep, the resonant energy is varied on a
time scale much longer than the average relaxation time of the
system~\cite{Williams2004a}.  Then entropy is conserved, and the system follows an
adiabatic path through the phase diagram.
The total entropy of the system, $S(T)$, is given by the sum of the entropies of each of the three components: 
\begin{eqnarray}
S_f(T,\mu_f) &=&  k_{\rm{B}}N_f \left[ 4 \frac{\mathcal F_4(z_f)}{\mathcal
F_3(z_f)} - \frac{\mu_f}{k_{\rm{B}}T} \right],
\\
S_b(T,\mu_b) &=& k_{\rm{B}}\tilde N_b \left[ 4 \frac{\mathcal G_4(z_b)}{\mathcal
G_3(z_b)} - \frac{\mu_b}{k_{\rm{B}}T} \right],
\\
S_{bf}(T,\mu_{bf},\epsilon_{\rm{res}}) &=& k_{\rm{B}}N_{bf} \left[ 4 \frac{\mathcal F_4(z_{bf})}{\mathcal
F_3(z_{bf})} - \frac{\mu_{bf}-\epsilon_{\rm{res}}}{k_{\rm{B}}T} \right].
\end{eqnarray}

The final temperature and chemical potentials
at the end of the sweep are determined from the condition $S(T^i) = S(T^f)$ together with the constraints (\ref{totN}) and (\ref{delNi}).
Figure~\ref{PhaseDiagram} shows the contours of constant entropy
in the phase diagrams for the case of equal populations.  Starting with a pure atomic system at $\epsilon_{\rm{res}}\gg k_{\rm{B}}T_{\rm{F}}$ we observe that there is a temperature, $T_{\rm{iso}}$, such that for $T^i
> T_{\rm{iso}}$ the final temperature after complete association of the gas is higher than the initial temperature. However, for $T^i<T_{\rm{iso}}$ we find the interesting
result that the temperature of the system decreases after the
atoms have been converted into molecules.  This can be understood
from the fact that at low temperatures, the majority of the free
bosons are condensed and do not contribute to $S$. By sweeping the value of $\epsilon_{\rm{res}}$
across the resonance we go from an almost perfect Fermi sea of atoms to
an almost perfect Fermi sea of molecules.  But since $\bar{\omega}_{bf}<\bar{\omega}_f$, the final temperature must be lower if  $S(T^i)=S(T^f)$.

An analytic expression for the relation between the initial and
final temperatures can be obtained in high- and low-$T$ limits. For $T \gg (T_{\rm{F}},T_c)$, the entropy of each component is of the Boltzmann form~\cite{Pathria}
\begin{equation}
S_k(T) = k_{\rm{B}}N_k \left\lbrace 4-\ln \left[ \left(
\frac{\hbar \bar{\omega}_k}{k_{\rm{B}}T}\right) N_k \right]\right
\rbrace.
\end{equation}
One obtains $T^f$ from $S(T^f)=S(T^i)$; in the case $x=1/2$ the result is
\begin{equation}
\frac{T^f}{T_{\rm{F}}}=6^{1/3} e^{4/3} \left( \frac{\bar{\omega}_{bf}}{\bar{\omega}_b}\right) \left( \frac{T^i}{T_{\rm{F}}} \right)^2.
\label{Tf_highT}
\end{equation}
The bosonic  entropy for $T\ll (T_{\rm{F}},T_c)$ is:
\begin{equation}
S_b(T) =  \frac{2\pi^4}{45} k_{\rm{B}} \left( \frac{k_{\rm{B}} T^i}{\hbar\bar{\omega}_b}
\right)^3, 
\end{equation}
while for the fermionic atoms the low temperature expansion of ${\mathcal{F}}_n(z)$~\cite{Pathria} yields:
\begin{equation}
S(T) = \pi^2k_{\rm{B}}N_f \left( \frac{T}{T_{\rm{F}}}
\right) \left[ 1-\frac{\pi^2}{5} \left(\frac{T}{T_{\rm{F}}}
\right)^2 \right].
\label{S_lowT_Fermi}
\end{equation}
The entropy of the molecules is given by a similar expression.
The adiabatic constraint $S(T^i)=S(T^f)$ in general does
not lead to a simple compact expression. However, by setting
$T^i=T^f\equiv T_{\rm{iso}}$ we find the initial temperature for
which an adiabatic sweep is also isothermal (in the sense that the
temperature endpoints are equal):
\begin{equation}
\frac{T_{\rm{iso}}}{T_{\rm{F}}} = \frac{\sqrt{15}}{\pi x^{1/3}}\sqrt{\frac{r_{bf}\left(1-x\right)^{2/3}+\left(2x-1\right)^{2/3}-x^{2/3}}{4 r_b^3+ 3 r_{bf}^3}}.
\label{Tisothermal}
\end{equation}
If $T^i>T_{\rm{iso}}$ the gas is heated by the adiabatic sweep, while a starting temperature $T^i<T_{\rm{iso}}$ leads to cooling.
Assuming extremely low temperatures, such that the entropy can be taken to be linear in $T$, a simple relation can be obtained for $\Delta N>0$:
\begin{equation}
\label{TfVSTi}
\frac{T^f}{T^i}=\frac{x^{2/3}}{(2x-1)^{2/3}+r_{bf}(1-x)^{2/3}},
\ \ \ x \geq 0.5.
\end{equation}
In the case of equal atomic populations, the relation becomes:
\begin{equation}
\label{TfvsTiLowTEqualN}
T^f= T^i /r_{bf} .
\end{equation}
The same expression holds in the case of $\Delta N<0$, where the ratio
$T^f/T^i$ is independent of $|\Delta N|$, as long as there is a condensate in both initial and final states. This is the regime where experiments are currently operating.

\begin{figure}
\centering
\includegraphics[width=\columnwidth]{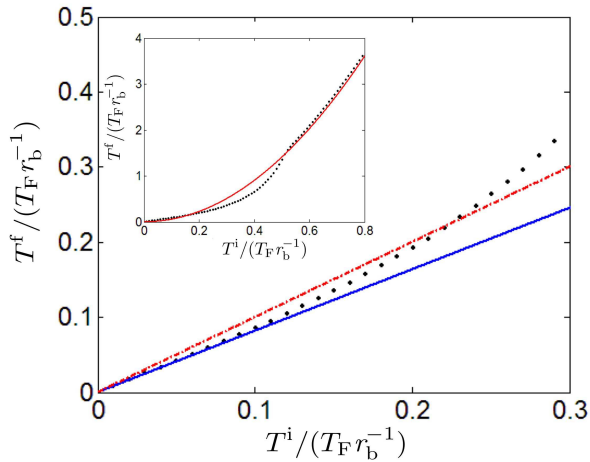}
\caption{Final temperature after adiabatic sweep from atomic to molecular configuration as a function of the initial temperature (dots). The frequency ratios correspond to those of the JILA experiment~\cite{JILA_parameters}, and $x=1/2$. The analytic relation (\ref{TfvsTiLowTEqualN}) is shown for comparison (solid line), along with the line $T^f=T^i$ (dashed line). The inset shows the same numerical results compared with the analytic expression (\ref{Tf_highT}) in the high temperature limit (solid line).} 
\label{Tfinal}
\end{figure}

Figure \ref{Tfinal} plots the dependence of the final temperature on the initial temperature for $x=1/2$ for the JILA trap parameters~\cite{JILA_parameters} in comparison to the low- and high-$T$ limits (\ref{TfvsTiLowTEqualN}) and (\ref{Tf_highT}), respectively. For these parameters Eq. (\ref{Tisothermal}) give $T_{\rm{iso}}=0.20T_{\rm{F}}$, which is seen to provide a good estimate of the value of $T^i$ for which $T^f=T^i$.

\begin{figure}
\centering
\includegraphics[width=\columnwidth]{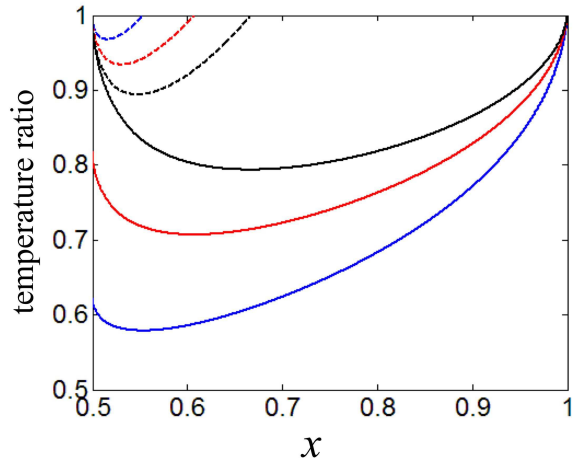}
\caption{Ratio $T^f/T^i$ after adiabatic conversion of atoms to molecules (Eq. (\ref{TfVSTi}); solid lines), and ratio $T^1/T^1_{\rm{F}}$ after one cooling cycle, steps 1-6, normalized by its initial value $T^0/T^0_{\rm{F}}$ (Eq. (\ref{TfTF_vs_n}); dashed lines), both as a function of the fermionic fraction $x$. The colors correspond to the JILA parameters~\cite{JILA_parameters} (red), the MIT parameters~\cite{MIT_parameters} (blue) and the best case $r_{bf}=1$ (black).}
\label{Tf_vs_x}
\end{figure}

{\bf{Cooling Cycle}}--We plot Eq.(\ref{TfVSTi}) in Fig.~\ref{Tf_vs_x}.  As is evident, $T^f/T^i$ depends strongly on the fermionic fraction $x$. In particular, there is an optimal fraction,
$x_{\rm{opt}}=[1+r_{bf}^3]/[1+2r_{bf}^3]$, which maximizes cooling; it corresponds to the minimum of the $T^f/T^i$ curve.

Since $x_{\rm{opt}}>1/2$ the following cooling cycle is possible:
\vspace{0.1cm}
\noindent 1. Start with an atomic gas at $\epsilon_{\rm{res}}/k_{\rm{B}}T_{\rm{F}} \gg 1$. Cool the gas to $T<T_{\rm{iso}}$
through standard cooling techniques.

\noindent2. Alter the fermionic fraction to $x_{\rm{opt}}$.

\noindent3. Sweep $\epsilon_{\rm{res}}$ adiabatically to $-\epsilon_{\rm{res}}/k_{\rm{B}}T_{\rm{F}} \gg 1$, forming the maximum number of molecules possible.

\noindent 4. Remove the free atoms from the system. 

\noindent 5. Adiabatically increase $\epsilon_{\rm{res}}$ into the dissociated regime.

\noindent 6. Continue the cycle by repeating steps 2 to 5 until the desired
final temperature in reached.
\vspace{0.1cm}

Assuming $T\ll T_{\rm{F}}$ we can use (\ref{TfVSTi}) and
(\ref{TfvsTiLowTEqualN}) to calculate the final temperature of
the system after $\textit{n}$ cycles: 
\begin{equation}
T^n = T^0 \left( \frac{r_{bf} x^{2/3}}{(2x-1)^{2/3} + r_{bf}(1-x)^{2/3}} \right)^n,
\end{equation}
where we have neglected a small decrease in temperature 
after suddenly decreasing the populations in steps 2 and 4. 
Notice that values of $r_{bf}$ closer to unity produce better cooling; this can be attained  by appropriate choice of wavelength of the trapping laser~\cite{SC}.

\begin{figure}[htbp]
\centering
\includegraphics[width=\columnwidth]{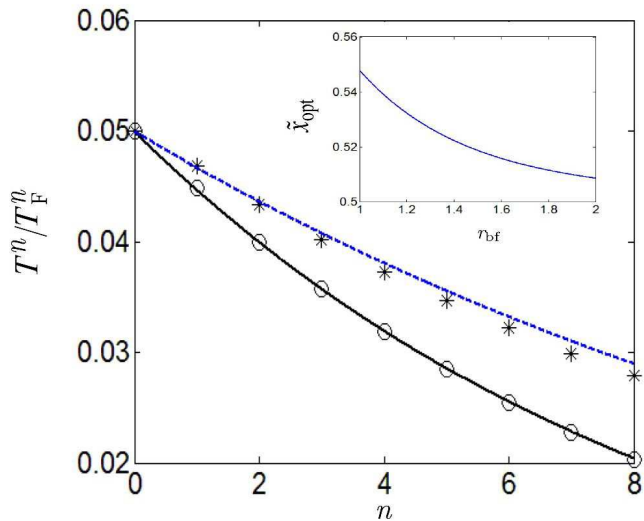}
\caption{Ratio of absolute to Fermi temperature after $n$ cycles operated at $x=\tilde{x}_{\rm{opt}}$. The numerical results (symbols) are compared with the analytical estimate (\ref{TfTF_vs_n}), which neglects a small temperature decrease that occurs in steps 2 and 4 (lines). Results are shown for the JILA parameters~\cite{JILA_parameters} (asterisks and dashed line) and for the best case $r_{bf}=1$ (circles and solid line). Inset: $\tilde{x}_{\rm{opt}}$ as a function of the frequency ratio $r_{bf}$.} 
\label{Tfinal_cycle}
\end{figure}

As is the case for evaporation, this cooling scheme leads to particle loss from the system in steps 2 and 4. This results in a reduction of the Fermi temperature $T_{\rm{F}}$.
Hence the figure of merit for increasing phase-space density is the decrease of the ratio of the temperature $T$ to the  Fermi temperature $T_{\rm{F}}$ {\textit{at the end of the cycle}}. 
A simple calculation shows that
after $\textit{n}$ cycles $T^n_{\rm{F}} = T_{\rm{F}}^0 [(1-x)/x]^{n/3}$.
The expression for $T/T_{\rm{F}}$ as a function of the number of cycles
performed is then:
\begin{equation}
\frac{T^n}{T^n_{\rm{F}}} = \frac{T^0}{T_{\rm{F}}^0} \left[ \frac{r_{bf} x}{(2x-1)^{2/3}(1-x)^{1/3}+r_{bf}(1-x)} \right]^n.
\label{TfTF_vs_n}
\end{equation}
Fig.~\ref{Tf_vs_x} shows this expression for $n=1$ and Fig.~\ref{Tfinal_cycle} shows an example of the change in $T/T_{\rm{F}}$,  as a function of the number of cycles performed.  
For $x>x_{\rm{opt}}$ the cycle results in a net increase of $T/T_{\rm{F}}$, and for $x=x_{\rm{opt}}$ the phase-space density remains unchanged. We find that there is a fermionic fraction, $\tilde{x}_{\rm{opt}}<x_{\rm{opt}}$, which maximizes the reduction of $T/T_{\rm{F}}$; this corresponds to the minimum of the respective curve depicted in Fig. \ref{Tf_vs_x}. We plot it as a function of $r_{bf}$ in the inset of Fig.~\ref{Tfinal_cycle}.

The efficacy of the proposed cooling scheme is contingent on losses from three-body collisions being minimal. Presently, no estimates of their rates exist for the systems under consideration. Additionally, phase separation may become an issue, as the interspecies atomic scattering length diverges on resonance~\cite{separation}. 

{\bf{Conclusion}}--We have presented the phase diagram of an ideal mixture of bosonic and fermionic atoms in thermal and chemical equilibrium with heteronuclear fermionic molecules. A cooling cycle has been identified that exploits the mechanism of adiabatic atom-molecule interconversion. 
This cycle could provide a useful complement to
existing cooling techniques.  Its practical efficiency will be limited by effects of atom-atom and atom-molecule interactions, which lie outside the scope of the present treatment, and must be determined by experiment.

MAM acknowledges support by a NIST SURF Summer Undergraduate Research Fellowship; NN and CWC acknowledge support by the National Science Foundation under grant PHY-0100767.

\end{document}